\documentclass[prd, twocolumn, floatfix, showkeys]{revtex4}
\usepackage{graphicx}
\usepackage{color}
\usepackage{amsmath, amsfonts, amssymb, bm}
\usepackage{slashed}
\begin{document}

\newcommand{\ka}{\slashed{k}}
\newcommand{\A}{\slashed{A}}
\newcommand{\p}{\slashed{p}}
\newcommand{\eps}{\slashed{\varepsilon}}

%%%%%%%%%%%%%%%%%%%%%%%%%%%%%%%%%%%%%%%%%%%%%%%%%%%%%%%%%%%%%%%%%
\title{Muon pair production in electron-positron collisions\\ 
close to threshold in the presence of a strong laser field}
\author{N.~Mahlin}
\author{S.~Villalba-Ch\'avez}
\author{C.~M\"uller}
\address{Institut f\"ur Theoretische Physik I, Heinrich-Heine-Universit\"at D\"usseldorf, Universit\"atsstra{\ss}e 1, 40225 D\"usseldorf, Germany}
\date{\today}
%%%%%%%%%%%%%%%%%%%%%%%%%%%%%%%%%%%%%%%%%%%%%%%%%%%%%%%%%%%%%%%%%
\begin{abstract}
Within the framework of strong-field quantum electrodynamics, we study the creation of muon-antimuon pairs in laser-assisted collisions of electrons with positrons, whose center-of-mass energy is close to the threshold of the process. Our focus lies on incident collision energies slightly below the threshold, where the associated energy gap has to be overcome by multiphoton absorption from the applied high-intensity laser field while the collision occurs. We calculate the cross section of the process and discuss its nonperturbative dependencies on the energy gap and the laser parameters. Three qualitatively different interaction regimes are identified, where the influence of the laser field either has a classical or fully quantum nature. In the latter case, an exponential dependence of the cross section on the collision parameters is found, which resembles the Schwinger effect.
\end{abstract}
%%%%%%%%%%%%%%%%%%%%%%%%%%%%%%%%%%%%%%%%%%%%%%%%%%%%%%%%%%%%%%%%%%%


\maketitle

\section{Introduction}
Creation of muon-antimuon pairs in the annihilating collision of electrons with positrons ($e^+e^-\to\mu^+\mu^-$) is one of the most basic particle reactions. The process possesses an energy threshold of $2Mc^2$ determined by the muon rest energy $Mc^2 \approx 105$\,MeV and has generally proven fundamental for our understanding of other $e^+e^-$ reactions \cite{Peskin}. While usually being mediated by an intermediate virtual photon, at high collision energies the reaction can also proceed via the exchange of a $Z^0$ boson \cite{electroweak}. In modern high-energy physics experiments at conventional accelerator facilities, $e^+e^-\to\mu^+\mu^-$ mainly represents a background process which, however, can be useful for calibration purposes. It is therefore important to know its cross section with high precision, including all kinds of radiative corrections \cite{radiative1, radiative2, radiative3}. Furthermore, $\mu^+\mu^-$ production in $e^+e^-$ collisions close to threshold has nowadays found renewed interest as potential muon source for a future muon collider \cite{LEMMA, near-threshold}.

In recent years, muon pair production has also been studied in the context of strong-field laser physics. Via wakefield acceleration in laser-generated plasmas, electrons (and in principle also positrons) can be accelerated to GeV energies. When such a high-energy electron beam impinges on a solid target foil, muon pairs can be produced in the Coulomb fields of the target nuclei \cite{Wakefield1, Wakefield2, Wakefield3}, as has recently been demonstrated in experiment \cite{Wakefield4, Wakefield5}. In this scheme, the laser field excerts an indirect impact by generating the incident GeV electron beam, but does not participate in the subsequent muon creation step. 

Theoreticians have, moreover, studied possibilities for $\mu^+\mu^-$ production with direct laser participation when the generating $e^+e^-$ collision occurs {\it inside} a laser field. The influence of a moderately intense background laser field on the reaction kinematics and the structure of the cross section of the process $e^+e^-\to\mu^+\mu^-$ has been analyzed in \cite{Roshchupkin}. A high-intensity laser wave, which is superposed collinearly to the $e^+e^-$ collision axis, can strongly enhance the available center-of-mass energy, because the field-dressed momentum of the copropagating particle largely exceeds its field-free momentum \cite{Boosted-McDonald, Boosted-PLB}. Particle reactions such as $\mu^+\mu^-$ production can also arise from coherent $e^+e^-$ collisions when a positronium atom is irradiated with high-intensity laser fields \cite{collider-EPL, Positronium-PRD, Positronium-PLB}. In this case, both particles are strongly laser-accelerated to gain a sufficiently high field-dressed collision energy. $\mu^+\mu^-$ production has also been considered in high-energy laser-driven re-collisions of $e^+e^-$ pairs that have been created from vacuum -- in a preceding reaction step -- by the impact of a relativistic nucleus \cite{Kuchiev} or a $\gamma$-photon \cite{Meuren} on a high-intensity laser wave. When the process $e^+e^-\to\mu^+\mu^-$ occurs at a very high center-of-mass energy in the $Z^0$ region, a background laser field of moderate intensity excerts only minor effects \cite{Ouali}.

In the present paper we study a complementary scenario of $\mu^+\mu^-$ production by laser-assisted $e^+e^-$ collisions. An $e^-$ and $e^+$ are assumed to impinge with a center-of-mass energy close to the threshold of $2Mc^2$, focussing in particular on energies slightly below the threshold. Muon pair production in the collision is enabled by the influence of a background laser field from which the missing energy can be absorbed to overcome the threshold. Our investigation is inspired by studies of field-induced $e^+e^-$ pair production via the dynamically assisted Schwinger effect \cite{Schwinger, Schutzhold-PRL2008, Lotstedt-PRL, DiPiazza-PRL2009, Orthaber-PLB2011, Akal-PRD2014, Otto-PLB2015, Otto-PRD2015, perturbative, Plunien-PRD2018, Taya, Selym-PRD2019, Kohlfurst, Mahlin-PRD}. In this situation, $e^+e^-$ pairs are generated in a slowly varying strong electric field and a superimposed rapidly oscillating weak field mode, whose photon energy lies below the threshold of $2mc^2$. By absorbing a photon from this assisting mode, the energy gap for pair production is therefore partially bridged, so that the low-frequency field only needs to overcome the remaining barrier. 

The laser-assisted $e^+e^-\to\mu^+\mu^-$ process studied in the present paper expands the research on dynamically assisted pair production from the first to the second lepton generation. In our case, the threshold energy is partially bridged by the collision energy of the incident $e^-$ and $e^+$, and the remaining barrier overcome by energy absorption from an applied background laser field. We will show that three qualitatively different interaction regimes exist which can be classified by the size of the energy gap and the intensity parameter of the laser field. While in two of the regimes the impact of the laser field has a classical character, the third regime exhibits a genuinely quantum, tunneling-like nature resembling the Schwinger effect, with a specific parameter dependence. 

From now on we use relativistic units with $\hbar=c=4\pi\varepsilon_0=1$, except where explicitly stated otherwise. Products of four-vectors are denoted as $(ab) = a_\mu b^\mu = a^0b^0 - \boldsymbol{a}\cdot\boldsymbol{b}$ and Feynman slash notation is applied.

\section{Theoretical framework}
We study the process $e^+e^-\to\mu^+\mu^-$ in the presence of an assisting laser field. The latter is assumed to be a monochromatic, plane wave of circular polarization with the classical four-potential 
\begin{eqnarray}
\label{A}
A^\mu(\varphi) = a\,[ \varepsilon_1^\mu\cos(\varphi) + \varepsilon_2^\mu\sin(\varphi) ],
\end{eqnarray}
where $a$ is the amplitude and $\varphi=(kx)$ the phase, with the wave four-vector $k^\mu=\omega(1,0,0,1)$. The polarization vectors satisfy $(k\varepsilon_i)=0$ and $(\varepsilon_i\varepsilon_j)=-\delta_{ij}$.

\begin{figure}[t]
\begin{center}
\includegraphics[width=0.36\textwidth]{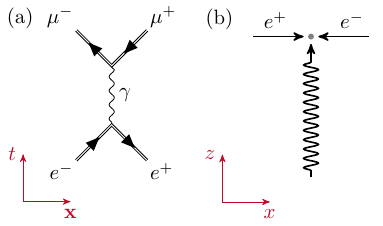}
\caption{(a) Feynman diagram of $e^+e^-\to\mu^+\mu^-$, mediated by a virtual photon, in the presence of a laser field. The double lines indicate the laser-dressed fermion states (Furry picture). (b) Scheme of the considered collision geometry.} 
\label{fig:diagram}
\end{center}
\end{figure}

Within the framework of laser-dressed QED, the amplitude for the laser-assisted process $e^+e^-\to\mu^+\mu^-$ reads 
\begin{eqnarray}
\label{S}
\mathcal{S} \!\!&=&\!\! {\rm i}\alpha\!\int\!{\rm d}^4x\int\!{\rm d}^4y\,\overline{\Psi}_{p_+,s_+}(x)\gamma^\mu\Psi_{p_-,s_-}(x)\,\mathcal{D}_{\mu\nu}(x-y) \nonumber\\
& &\qquad\qquad\quad\,\times\,\overline{\Psi}_{P_-,S_-}(y)\gamma^\nu\Psi_{P_+,S_+}(y)\,,
\end{eqnarray}
where $\alpha$ denotes the fine-structure constant and
\begin{eqnarray}
\mathcal{D}_{\mu\nu}(x-y) = \int \frac{{\rm d}^4q}{(2\pi)^4}
                  \frac{4\pi{\rm e}^{{\rm i}q\cdot(x-y)}}{q^2+{\rm i}0^+}\,g_{\mu\nu}
\end{eqnarray}
is the photon propagator in the Feynman gauge; see Fig.~\ref{fig:diagram}(a). The laser-dressed states for the electron and positron are given by \cite{Landau}
\begin{eqnarray}
\label{Vol}
\Psi_{p_\pm,s_\pm}(x) = \sqrt{\frac{m}{q_\pm^0V}} \left(1\pm\frac{e\ka\A}{2(k{p_\pm})}\right) u_{p_\pm,s_\pm} {\rm e}^{{\rm i}f^{\scriptscriptstyle{(\pm)}}}
\end{eqnarray}
with
\begin{eqnarray*}
f^{\scriptscriptstyle{(\pm)}} = 
\pm (q_\pm x) + \frac{ea(p_\pm\varepsilon_1)}{(kp_\pm)}\sin(kx) - \frac{ea(p_\pm\varepsilon_2)}{(kp_\pm)}\cos(kx).
\end{eqnarray*}
In Eq.\,(\ref{Vol}), $p_\pm^\mu$ are the initial free four-momenta of the electron and positron 
(outside the laser field), $s_\pm$ denote the particle spin states, the $u_{p_\pm,s_\pm}$ are 
free Dirac spinors, and $q_\pm^\mu = p_\pm^\mu + \frac{e^2a^2}{2(kp_\pm)}k^\mu$ the effective four-momenta of the particles in the laser field, with $e$ denoting the positrons charge and $V$ a normalization volume. The corresponding effective mass $m_*$ follows from
$m_*^2 = q_\pm^2 = (1 + \xi^2)m^2$ with the dimensionless laser intensity parameter
$\xi = \frac{ea}{m}$. Analogous expressions hold for the Volkov states $\Psi_{P_\pm,S_\pm}$,
the free momenta $P^\mu_\pm$, the spin states $S_\pm$, the effective momenta $Q_\pm^\mu$, the effective mass $M_\ast=M(1+\Xi^2)^{1/2}$, and the intensity parameter $\Xi=\frac{ea}{M}$ of the muons.

By the standard procedure of using the generating function of the Bessel functions \cite{AS}, one can perform the space-time integrations in Eq.\,(\ref{S}) to get
\begin{eqnarray}
\label{S1}
\mathcal{S}\!\! &=&\!\! \frac{{\rm i}\alpha}{V^2} \frac{mM}{\sqrt{q_+^0q_-^0Q_+^0Q_-^0}} \frac{(2\pi)^6}{\pi}
\int \frac{{\rm d}^4q}{q^2+{\rm i}0^+}\nonumber\\
& & \!\times\sum_{n,N} \mathcal{M}^\mu(p_+,p_-|n) \mathcal{M}_\mu(P_-,P_+|N) \nonumber\\
& & \!\times\,\delta(q_+ + q_- -q + nk)\, \delta(Q_+ + Q_- -q -Nk)
\end{eqnarray}
with spinor-matrix products \cite{Positronium-PRD}
\begin{eqnarray}
&&\mathcal{M}^\mu(p_+,p_-|n)=\overline{u}_{p_+,s_+}\,\widetilde{m}_n^\mu\,u_{p_-,s_-}\ ,\nonumber\\
&&\mathcal{M}^\mu(P_-,P_+|N)=\overline{u}_{P_-,S_-}\,\widetilde{M}_N^\mu\,u_{P_+,S_+}
\end{eqnarray}
and matrices
\begin{eqnarray}
\widetilde{m}_n^\mu\!\! &=&\!\! b_n\!\left(\gamma^\mu-\frac{e^2a^2}{2}\frac{k^\mu\ka}{(kp_+)(kp_-)}\right)\nonumber\\
& & \!+\big(c_n\varepsilon_1^\nu+d_n\varepsilon_2^\nu\big)\frac{ea}{2}\!\left[\frac{\gamma_\nu\ka\gamma^\mu}{(kp_+)}-\frac{\gamma^\mu\ka\gamma_\nu}{(kp_-)}\right],\nonumber\\
\widetilde{M}_N^\mu\!\! &=&\!\! B_N\!\left(\gamma^\mu-\frac{e^2a^2}{2}\frac{k^\mu\ka}{(kP_-)(kP_+)}\right)\nonumber\\
& & \!-\big(C_N\varepsilon_1^\nu+D_N\varepsilon_2^\nu\big)\frac{ea}{2}\!\left[\frac{\gamma_\nu\ka\gamma^\mu}{(kP_-)}-\frac{\gamma^\mu\ka\gamma_\nu}{(kP_+)}\right].
\end{eqnarray}
The coefficients in these matrices are $b_n=J_n(z)\,{\rm e}^{{\rm i}n\phi}$, $c_n=\frac{1}{2}\left(b_{n+1}+b_{n-1}\right)$, and $d_n=\frac{1}{2i}\left(b_{n+1}-b_{n-1}\right)$ at the electronic vertex as well as $B_N=J_{-N}(Z)\,{\rm e}^{{\rm i}N\Phi}$, $C_N=\frac{1}{2}\left(B_{N+1}+B_{N-1}\right)$, and $D_N=\frac{1}{2i}\left(B_{N+1}-B_{N-1}\right)$ at the muonic vertex. Here, the Bessel function arguments and phases of the exponentials are given by
$z=\sqrt{z_1^2+z_2^2}$, $\cos(\phi)=z_1/z$, $\sin(\phi)=z_2/z$, and 
$z_j=ea(\varepsilon_jp_+)/(kp_+)-ea(\varepsilon_jp_-)/(kp_-)$ for $j\in\lbrace1,2\rbrace$. Analogous formulas hold for $Z$, $\Phi$, and $Z_j$, which arise from the above expressions by replacing the momenta $p_\pm^\mu\to P_\pm^\mu$.

As is expressed by the energy-momentum conserving $\delta$-function at the first vertex, the integer number $n$ in Eq.\,(\ref{S1}) counts the laser photons that are absorbed (if $n>0$) or emitted (if $n<0$) by the electron and positron. Similarly, $N$ is the number of absorbed or emitted laser photons at the muon vertex. Integrating over the virtual photon momentum and applying Graf's addition theorem for Bessel functions \cite{AS}, the amplitude adopts the structure 
\begin{eqnarray}
\label{S2}
\mathcal{S}\!\! &=&\!\!\frac{{\rm i}\alpha}{V^2} \frac{mM}{\sqrt{q_+^0q_-^0Q_+^0Q_-^0}} \frac{(2\pi)^6}{\pi} \frac{1}{\left(q_++q_-\right)^2}\nonumber\\
& & \!\times\sum_{\mathcal{N}}\delta(Q_++Q_--q_+-q_--\mathcal{N}k)\,T_{_\mathcal{N}}\,,
\end{eqnarray}
where $\mathcal{N}\equiv n+N$ is the total number of laser photons, which are absorbed from or emitted into the laser field, and $T_{_\mathcal{N}}$ are rather complicated functions of the particle momenta and laser parameters. Note that in Eq.~\eqref{S2} we have used the approximation $q^2 = (q_+ + q_- + nk)^2 \approx (q_+ + q_-)^2$ since $n\omega \sim m \ll q_\pm^0 \approx M_\ast$ holds for the collision parameters of interest here (see Sec.~III). In connection, we have also dropped the ${\rm i}0^+$ from the photon propagator, since the squared four-momentum of the exchanged photon $q^2 \approx 4M_*^2$ is far away from zero in the considered collisions.

The total cross section for laser-assisted $e^+e^-\to\mu^+\mu^-$ is found from the square of 
the amplitude \eqref{S2}, averaged and summed over the particle spins, 
integrated over the outgoing muon momenta, and divided by the interaction time and the incoming leptons current density $j_q=I_q/(Vq_+^0q_-^0)$, with $I_q=\sqrt{(q_+q_-)^2-m_\ast^4}$:
\begin{eqnarray}
\label{sigma}
\sigma = \frac{V^2}{j_qT}\int\frac{{\rm d}^3Q_+}{(2\pi)^3}\int\frac{{\rm d}^3Q_-}{(2\pi)^3}\,\frac{1}{4}\sum_{s_\pm,S_\pm}\lvert \mathcal{S}\rvert^2\,.
\end{eqnarray}

\section{Results and Discussion}

Within the formalism of laser-dressed QED described in Sec.~II, we have calculated the process $e^+e^-\to\mu^+\mu^-$ in an intense laser field of circular polarization. Our focus lies on scenarios where the collision energy $\sqrt{s_p} = \sqrt{(p_++p_-)^2}$ of the incident electron and positron lies slightly below the process threshold $2M$. The corresponding energy gap is denoted as $\Delta_p := 2M-\sqrt{s_p}$; it has to be overcome by the interaction of the particles with the background laser field. We shall typically assume $\Delta_p$ to be of order $m$, so that its size is comparable to the energy gap of laser-induced $e^+e^-$ pair creation processes \cite{Ritus-Review}.    

\begin{figure}[b]
\begin{center}
\includegraphics[width=0.5\textwidth]{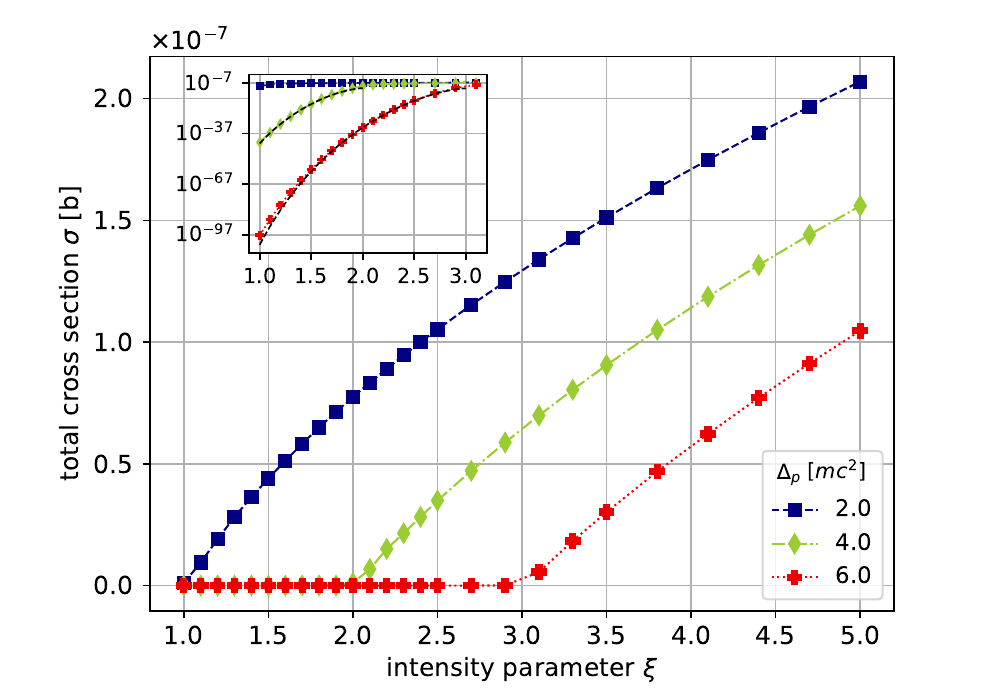}
\caption{Total cross section of subthreshold muon pair production, as function of the intensity parameter of the laser, for three different values of the energy gap $\Delta_p=2M-\sqrt{s_p}$, as indicated, and a laser frequency of $\omega=0.05\,m$. The inset shows the region $1.0\le \xi\le 3.1$ on a logarithmic scale, with dashed black curves depicting the analytical function \eqref{sigma-Schwinger-like}.}
\label{fig:sigma_xi}
\end{center}
\end{figure}

To be specific, we assume the collision geometry illustrated in Fig.~\ref{fig:diagram}(b). The electron and positron collide head-on along the $x$-axis and have equal free energies $p_+^0=p_-^0$. The laser wave impinges under 90$^\circ$, so that collision occurs within the polarization plane. Throughout most of our numerical calculations, we shall assume values $\xi\sim 1$ for the laser intensity parameter and take $\omega=0.05m$ for the laser frequency. We note that the latter value, which is chosen to facilitate the computations, lies in the x-ray domain, and is thus much larger than the frequency of typical high-power optical lasers. We will come back to this point at the end of Sec.~III.

Figure~\ref{fig:sigma_xi} shows the total $e^+e^-\to\mu^+\mu^-$ cross section inside a laser field for energy gaps of $2m$, $4m$, and $6m$, respectively, when the intensity parameter is varied between $1\le\xi\le 5$. Note that the applied laser frequency always satisfies $\omega\ll\Delta_p$. Since the collision energy $\sqrt{s_p}$ lies below the threshold for muon pair production, the process is rendered possible only through the presence of the laser field. The figure shows that sizeable cross sections $\sigma\sim 10^{-7}$\,b are attained, provided the laser intensity is large enough. (Recall that the cross section for $e^+e^-\to\mu^+\mu^-$ in vacuum attains a maximum value of $\approx 10^{-6}$\,b \cite{Peskin}.) The intensity required to reach sizeable cross sections grows with the size of $\Delta_p$. For example, for $\Delta_p=4m$ ($\Delta_p=6m$), the cross section starts to grow substantially at $\xi\approx 2$ ($\xi\approx 3$). When instead the laser intensity is too small, the cross sections fall off very quickly. Hence, a transition between different interactions regimes occurs when the laser intensity is varied.

\begin{figure}[t]
\begin{center}
\includegraphics[width=0.5\textwidth]{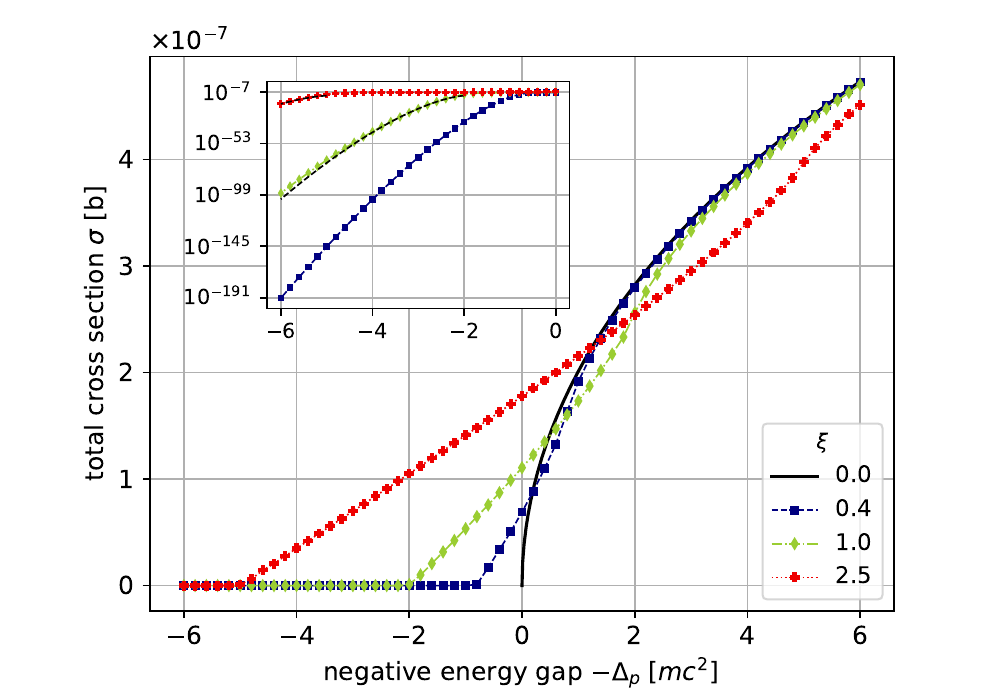}
\caption{Total cross section of muon pair production in a laser field, as function of the negative energy gap $-\Delta_p=\sqrt{s_p}-2M$ for three different values of the intensity parameter $\xi$, as indicated, and a laser frequency of $\omega=0.05\,m$. For comparison, the solid black line shows the total cross section in vacuum. The inset shows the region $-\Delta_p \le 0$ on a logarithmic scale, with the analytical curve \eqref{sigma-Schwinger-like} drawn in dashed black style.}
\label{fig:sigma_Dp}
\end{center}
\end{figure}

This interpretation is corroborated by Fig.~\ref{fig:sigma_Dp}, which shows the total $e^+e^-\to\mu^+\mu^-$ cross section for different values of $\xi$, when the energy gap is varied between $-6m\le \Delta_p\le 6m$, covering collision energies below and above the process threshold. Note that it is convenient here to show the graphs as function of the {\it negative} energy gap; this way, the collision energy increases from left to right. In the subthreshold region ($-\Delta_p<0$), we see again a characteristic transition in the size and shape of the cross sections. It occurs at $\Delta_p\approx 5m$ for $\xi=2.5$, at $\Delta_p\approx 2m$ for $\xi=1$, and at $\Delta_p\approx 0.8m$ for $\xi=0.4$. Interestingly, yet another transition occurs in the region above the threshold ($-\Delta_p>0$), where the curves noticeably change their slopes. For example, for $\xi=1$ ($\xi=2.5$) this happens at about $\Delta_p\approx -2m$ ($\Delta_p\approx -5m$). After having passed these points, the curves closely approach the vacuum cross section (shown by the solid black line). Before, however, the cross sections inside the laser field are smaller than the vacuum value. This is intriguing, because it means that in this region of above-threshold collision energies (where no laser field would be needed to enable the process) the presence of the laser field hampers the muon pair production.

Accordingly, there are three different interaction regimes in the process of $e^+e^-\to\mu^+\mu^-$ in a laser field. The physical origin of these regimes can be explained within an intuitive semiclassical picture. The classical four-momenta of the positron and electron inside the laser field are given by \cite{Ritus-Review}
\begin{eqnarray}
p_\pm^\mu(\varphi) = p_\pm^\mu \mp e A^\mu(\varphi) + \frac{m^2\xi^2}{2\omega p_\pm^0}k^\mu \mp \frac{m\xi p_{\pm,x}}{\omega p_\pm^0} \cos\varphi\, k^\mu. \nonumber\\
\end{eqnarray}
The combination of the first and third terms in this expression represents the effective (i.e.~phase-averaged) momenta $q_\pm^\mu$ of the particles. The last term, which is linear in the applied laser field, has the important property to coincide for the electron and the positron. This is because both particles have opposite charges as well as opposite momenta along the $x$-axis, leading to compensation of these two signs. The instantaneous, phase-modulated collision energy amounts to 
\begin{eqnarray}
\label{s_phi}
\sqrt{s(\varphi)} = \big(s_p + 4m^2\xi^2 - 8m\xi p_{+,x} \cos\varphi\big)^{1/2}.
\end{eqnarray}
The first two terms inside the square root coincide with $s_q = (q_+ + q_-)^2 = s_p + 4m^2\xi^2$, which represents the squared effective collision energy. We note that the threshold condition $s_p\ge 4M^2$ is equivalent to $s_q\ge 4M_*^2 = 4M^2 + 4m^2\xi^2$. For the parameters under consideration, where $2m\xi\ll\sqrt{s_p} = 2M - \Delta_p$, we can write approximately
\begin{equation}
\label{s_phi_approx}
\sqrt{s(\varphi)}\approx 2M - \Delta_p - 2m\xi\cos\varphi\,.
\end{equation}

\begin{figure}[b]
\begin{center}
\includegraphics[width=0.5\textwidth]{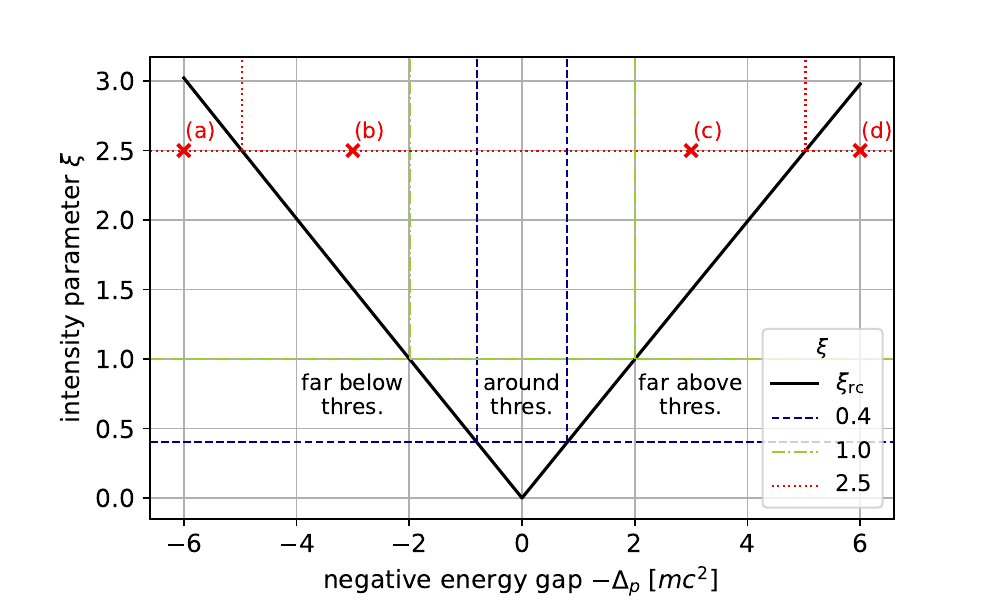}
\caption{Locations of the {\it far-below threshold}, {\it around threshold}, and {\it far-above threshold} interaction regimes in the $(\xi,-\Delta_p)$ plane. The regime changes occur at $\xi=\xi_{\rm rc}\approx \frac{|-\Delta_p|}{2m}$ (solid black line). The energy gaps, where the regime changes arise for the laser intensity parameters $\xi \in \{0.4, 1.0, 2.5\}$ from Fig.~\ref{fig:sigma_Dp}, are highlighted with dashed blue, dash-dotted green, and dotted red lines, respectively. Red crosses, moreover, mark the parameter combinations $(\xi,-\Delta_p)$ for which the partial cross sections in Fig.~\ref{fig:sigma_N} are given.}
\label{fig:Phase_diagram}
\end{center}
\end{figure}

This phase-modulated classical collision energy allows us to distinguish the following three qualitatively distinct interaction regimes:\\
(i) {\it far-below threshold regime}, where $\Delta_p > 2m\xi$. Here, $\sqrt{s(\varphi)} < 2M$ holds for all laser phases $\varphi$. The process possesses a tunneling-like, exponential characteristic, similar to the Schwinger effect [see Eq.~\eqref{sigma-Schwinger-like} below];\\
(ii) {\it around threshold regime}, where $|\Delta_p| < 2m\xi$. In this intermediate case, $\sqrt{s(\varphi)}$ is sometimes larger and sometimes smaller than $2M$, depending on the value of $\varphi$;\\
(iii) {\it far-above threshold regime}, where $-\Delta_p > 2m\xi$. Here, $\sqrt{s(\varphi)} > 2M$ for all laser phases $\varphi$. The process has a laser-assisted nature and its total cross section reaches the field-free value. 

This physically intuitive classification of the interaction regimes is illustrated in Fig.~\ref{fig:Phase_diagram} in the plane spanned by the negative energy gap $-\Delta_p$ and the intensity parameter $\xi$. It is in perfect accordance with the results in Figs.~\ref{fig:sigma_xi} and \ref{fig:sigma_Dp}. The transition between the far-below and around threshold regimes has occurred there at the positions where $0 < \Delta_p \approx 2m\xi$ holds, while the transition to the far-above threshold regime was located at $0 < -\Delta_p \approx 2m\xi$. 

By analyzing our results we have found that the cross section in the far-below threshold regime for $\xi\gtrsim 1$ can be well approximated by a function of the form
\begin{eqnarray}
\label{sigma-Schwinger-like}
\sigma_{\rm fbt}\approx \sigma_{\rm fbt}^{(0)}\,\exp\!\left(-\frac{4}{3}\frac{(\Delta_p-2m\xi)^{3/2}}{\omega\sqrt{m\xi}}\right),
\end{eqnarray}
whose parameter dependencies resemble the famous Schwinger rate $\sim\exp(-\pi \mathcal{E}_c/\mathcal{E}_0)$ of $e^+e^-$ pair production in a constant electric field $\mathcal{E}_0$, where $\mathcal{E}_c=m^2/e$ is the critical field strength of QED \cite{Schwinger}. The Schwinger rate describes a tunneling process through a barrier of height $2m$. Similar non-analytic field dependencies are also known from other pair creation processes in very strong and slowly oscillating fields (such as the nonlinear Breit-Wheeler effect \cite{Ritus-Review, Mahlin-PRD}) as well as from strong-field ionization of atoms whose rate scales as $\sim\exp\big(-\frac{4\sqrt{2m}}{3e\mathcal{E}_0}E_a^{3/2}\big)$ \cite{Landau-QM}. The atomic binding potential $E_a$ enters here in the same way as the field-modulated energy gap $\Delta_p-2m\xi$ in Eq.~\eqref{sigma-Schwinger-like}. The modulation of the collision energy by the field thus exerts a similar effect as a superimposed high-frequency field mode in the dynamically assisted Schwinger mechanism, which softens the strong exponential suppression and can accordingly lead to a large amplification of the process \cite{Schutzhold-PRL2008, Lotstedt-PRL, DiPiazza-PRL2009, Orthaber-PLB2011, Akal-PRD2014, Otto-PLB2015, Otto-PRD2015, perturbative, Plunien-PRD2018, Taya, Selym-PRD2019, Kohlfurst, Mahlin-PRD}.

As the insets in Figs.~\ref{fig:sigma_xi} and \ref{fig:sigma_Dp} illustrate, the analytical formula \eqref{sigma-Schwinger-like} provides a good approximation to the cross section. The pre-exponential factor $\sigma_{\rm fbt}^{(0)}$ has been treated as a fit parameter, which depends on $\xi$, $\omega$, and $\Delta_p$; for the scenarios displayed in the figures its optimized value lies between about 4.6 and $6.7\times 10^{-11}$\,b. Note in connection that the analytical curve \eqref{sigma-Schwinger-like} is not shown for the $\Delta_p=2m$ data in Fig.~\ref{fig:sigma_xi}, because the far-below threshold regime lies outside the plot range in this case. Likewise, the Eq.~\eqref{sigma-Schwinger-like} is not shown for the $\xi=0.4$ data in Fig.~\ref{fig:sigma_Dp} either, as it does not apply to this small $\xi$ value.

\begin{figure}[t]
\begin{center}
\includegraphics[width=0.5\textwidth]{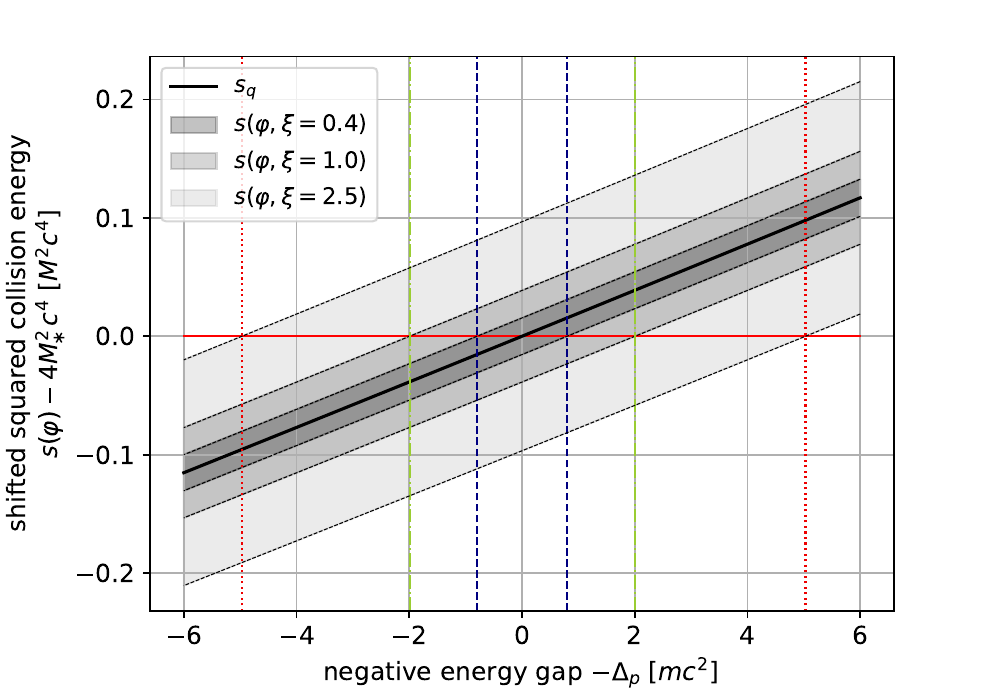}
\caption{Squared effective collision energy $s_q$ (solid black line) and squared field-modulated collision energies $s(\varphi)$ for different laser intensities (shaded regions), as function of the negative energy gap $-\Delta_p=\sqrt{s_p}-2M$. For graphical convenience, the ordinate is shifted by $4M_\ast^2$, as indicated by the solid red line. The energy gaps, where regime changes occur, are marked for the intensity parameters $\xi\in\{0.4, 1.0, 2.5\}$ from Fig.~\ref{fig:sigma_Dp} with vertical dashed blue, dash-dotted green, and dotted red lines, respectively.}
\label{fig:Energy_diagram}
\end{center}
\end{figure}

The field-modulated collision energy also allows us to explain, why the cross section in the laser field can lie sizeably below its field-free value, when $0 < -\Delta_p < 2m\xi$ (see Fig.~\ref{fig:sigma_Dp}). While the collision energy in this case is (slightly) above the threshold, the electron and positron are driven by the electromagnetic forces of the laser field, which accelerates or decelerates their motion, depending on the laser phase. When the deceleration is too strong, the instantaneous collision energy falls below the threshold, so that $\mu^+\mu^-$ pairs cannot be produced efficiently (i.e.~without additional absorption of laser photons) in the corresponding phase range, which reduces the total cross section. The field modulation of the collision energy is illustrated in Fig.~\ref{fig:Energy_diagram}. 

\begin{figure}[t]
\begin{center}
\includegraphics[width=0.45\textwidth]{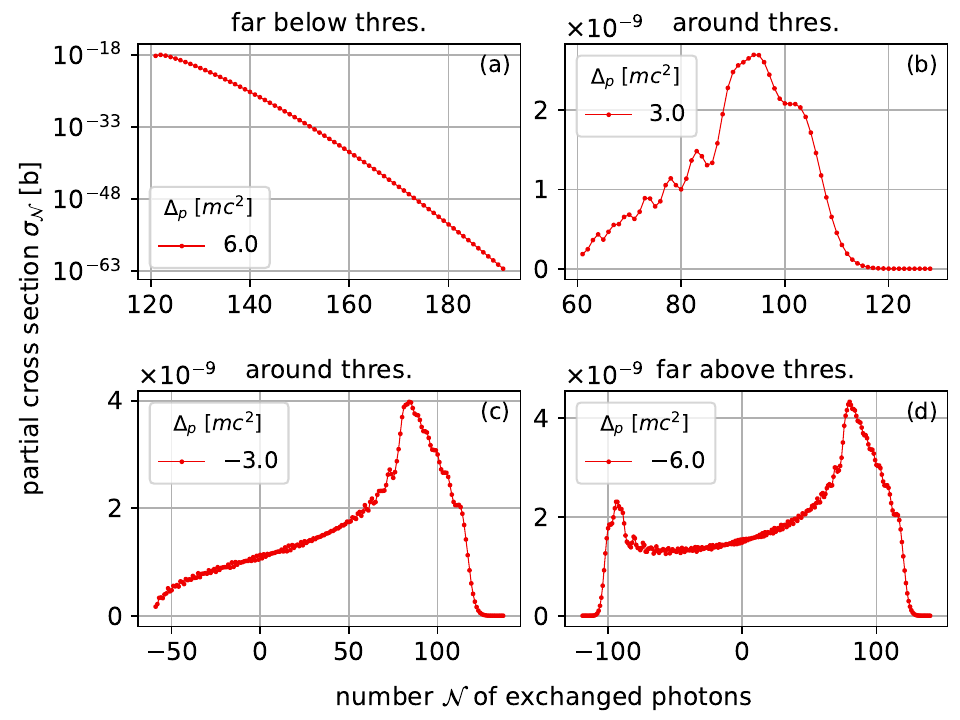}
\caption{Partial cross sections of muon pair production, as function of the total number of exchanged photons $\mathcal{N}$ for different energy gaps, spanning the regimes far below threshold (a), around threshold (b) and (c) and far above threshold (d). Data is given for a laser frequency of $\omega=0.05\,m$ and the laser intensity parameter $\xi=2.5$.}
\label{fig:sigma_N}
\end{center}
\end{figure}

The different characteristics of the three interaction regimes are also prominent in the distributions of the partial cross sections $\sigma_{_{\!\mathcal{N}}}$ in Fig.~\ref{fig:sigma_N}, which refer to a certain number $\mathcal{N}$ of photons exchanged with the laser field and build the total cross section by virtue of $\sigma = \sum_\mathcal{N}\sigma_{_{\!\mathcal{N}}}$. From panel (a) to (d), successively decreasing energy gaps of $\Delta_p=6m$, $3m$, $-3m$, and $-6m$ are assumed, this way going from the far-below threshold regime [Fig.~\ref{fig:sigma_N}(a)] over the around-threshold regime [Figs.~\ref{fig:sigma_N}(b) and (c)] to the far-above threshold regime [Fig.~\ref{fig:sigma_N}(d)]. 

In the scenario of Fig.~\ref{fig:sigma_N}(a), the largest $\sigma_{_{\!\mathcal{N}}}$ values are attained near the minimal number of laser photons $\mathcal{N}_{\rm min}=\frac{\Delta_p}{\omega}+1=121$ that need to be absorbed to overcome the energy gap. For larger $\mathcal{N}$, the partial cross sections fall very steeply in an exponential manner. Note that this figure is shown on a logarithmic scale, in contrast to the remaining panels. When the energy gap is reduced to $3m$ in Fig.~\ref{fig:sigma_N}(b), the shape of the $\sigma_{_{\!\mathcal{N}}}$ distribution changes drastically. Starting from the contribution for the smallest possible $\mathcal{N}$, the partial cross sections grow in a non-monotonous way until they reach a maximum value around $\mathcal{N}\approx 95$, from where on the contributions decrease in a step-like manner. 

The same kind of behaviour of the partial cross sections arises for an energy gap of $-3m$ in Fig.~\ref{fig:sigma_N}(c), where it is, however, more pronounced. Since the energy gap is negative here, the process may involve the emission of laser photons, so that the distribution starts at $\mathcal{N}_{\rm min}=-59$. From there it grows towards a maximum and afterwards shows a distinct step-like decline. When the process occurs at $\Delta_p=-6m$ far above the threshold in Fig.~\ref{fig:sigma_N}(d), a second maximum emerges shortly after the $\mathcal{N}_{\rm min}$ region. In this regime, the total cross section, which is obtained by summing over all $\sigma_{_{\!\mathcal{N}}}$, closely approaches the field-free value (see Fig.~\ref{fig:sigma_Dp}). The effect of the laser is to redistribute the energies and momenta of the created muons and antimuons, since each $\mathcal{N}$ corresponds to a slightly different collision energy. 

When looking at the panels (d) to (a) of Fig.~\ref{fig:sigma_N} in backward order, one can approximately recognize that a smaller and smaller portion of the distribution in (d) is cut when the energy gap is increased: to begin with, the second maximum is lost in (c); then in addition a substantial portion of the growing flank disappears in (b); and finally only the decreasing flank right from the main maximum is left in (a). The locations of the parameter combinations underlying panels (a)-(d) in the $(\xi,-\Delta_p)$ plane are indicated in Fig.~\ref{fig:Phase_diagram}. 

\begin{figure}[t]
\begin{center}
\includegraphics[width=0.5\textwidth]{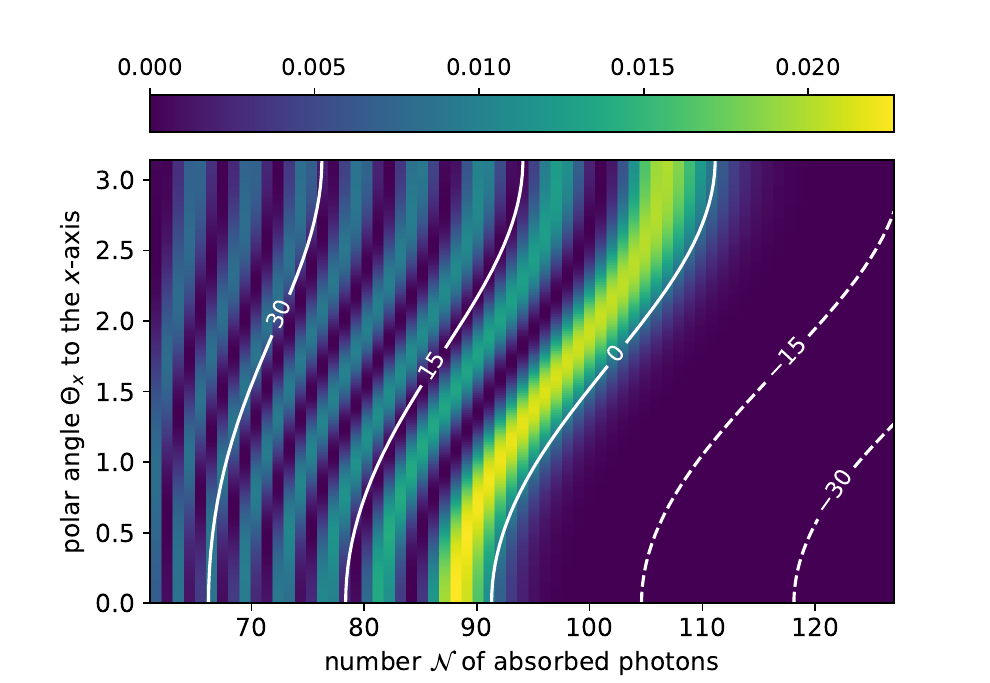}
\caption{Dependence of the squared Bessel function $J_\mathcal{N}^2(\zeta)$ on the total number of absorbed photons $\mathcal{N}$ and the polar emission angle $\Theta_x$ of the created antimuons effective momentum $\boldsymbol{Q}_+$, for a fixed azimuthal angle $\Phi_x=0$. The laser frequency $\omega=0.05\,m$, intensity parameter $\xi=2.5$, and energy gap $\Delta_p=3m$ are chosen as in Fig.~\ref{fig:sigma_N}(b). The white contour lines give the quantity $\zeta-\mathcal{N}$.}
\label{fig:Bessel_squared}
\end{center}
\end{figure}

In the considered parameter regime, where $|\Delta_p|,~m\xi,~\mathcal{N}\omega\ll M$ holds, we have found that the partial cross sections may be approximated as
\begin{eqnarray}
\label{sigma-approx}
\sigma_{_{\!\mathcal{N}}} \approx \frac{\alpha^2}{6} \frac{s_{_\mathcal{N}}(s_{_\mathcal{N}}+2M^2)}{s_q^2 I_q} \sqrt{1-\frac{4M_*^2}{s_{_\mathcal{N}}}} \int J_\mathcal{N}^2(\zeta)\,{\rm d}\Omega
\end{eqnarray}
with $s_{_\mathcal{N}}=(q_+ + q_- + \mathcal{N}k)^2$, $I_q$ given above Eq.~\eqref{sigma}, and the Bessel argument $\zeta = ea\sqrt{-(ww)}$ for $w^\mu = \frac{Q_+^\mu}{(kQ_+)}-\frac{Q_-^\mu}{(kQ_-)}-\frac{q_+^\mu}{(kq_+)}+\frac{q_-^\mu}{(kq_-)}$. Since $\sigma_{_{\!\mathcal{N}}}$ is a Lorentz-invariant quantity, it may be evaluated in any frame of reference; for given $\mathcal{N}$, we take the center-of-mass frame of the created muon pair. The solid angle element ${\rm d}\Omega = \sin(\Theta_x)\,{\rm d}\Theta_x\,{\rm d}\Phi_x$ for the angles of the antimuons momentum $\boldsymbol{Q}_+$ in this frame is defined with respect to the $x$-axis, with $\Phi_x$ measuring the angle to the $y$-axis.  

The dependence of $J_\mathcal{N}^2(\zeta)$ on the polar angle of the created antimuon is shown in Fig.~\ref{fig:Bessel_squared}. Largest values are reached when the argument $\zeta$ lies close to the order $\mathcal{N}$ \cite{AS}. By taking also the factor $\sin(\Theta_x)$ from the solid angle element into consideration, we can infer from Fig.~\ref{fig:Bessel_squared} the emergence of a maximum partial cross section at about $\mathcal{N}\approx 95$ in Fig.~\ref{fig:sigma_N}(b) and a subsequent step-wise decline of the $\sigma_{_{\!\mathcal{N}}}$ values when $\mathcal{N}$ increases further. 

A Bessel-function analysis of the far-below threshold regime enables us to explain a particular feature of the cross section in Eq.~\eqref{sigma-Schwinger-like}, whose field dependence does not coincide exactly with the Schwinger rate. It contains a square root of the field strength in the denominator of the exponent, accompanied by a square root of the field frequency (note that $\omega\sqrt{m\xi} = \sqrt{\omega e\mathcal{E}_0}$). The parameter dependence in Eq.~\eqref{sigma-Schwinger-like} follows from a Bessel function of the asymptotic form $J_\mathcal{N}(\zeta) \sim \exp\left(-\frac{2\sqrt{2}}{3}\frac{(\mathcal{N}-\zeta)^{3/2}}{\zeta^{1/2}}\right)$ \cite{AS} that enters in Eqs.~\eqref{S2} and \eqref{sigma-approx}. While this is in principle typical for a tunneling phenomenon, the unusual feature of the present process is that the main contribution to the sum in Eq.~\eqref{S2} stems from the smallest possible photon numbers $\mathcal{N}\approx \mathcal{N}_{\rm min} \approx \Delta_p/\omega$ [see Fig.~\ref{fig:sigma_N}(a)]. This is because the Bessel function argument $\zeta$ is mainly determined by the incident electron and positron momenta (with $p_\pm^0\gg \mathcal{N}\omega$) and therefore depends only very weakly on $\mathcal{N}$. Hence, the difference $\mathcal{N}\!-\!\zeta$ is minimized and thus the value of $J_\mathcal{N}(\zeta)$ maximized for $\mathcal{N}\approx\mathcal{N}_{\rm min}$. In contrast, in other Schwinger-like processes, the dominant contributions arise from photon numbers largely exceeding the smallest kinematically allowed value \cite{Ritus-Review}.

Before moving on to the conclusion we comment on the field parameters used in our numerical calculations. Laser pulses with intensity parameters $\xi\gtrsim 1$ are routinely generated at high-field facilities. The associated laser frequencies typically lie in the optical range $\omega\sim 1$\,eV, whereas we have applied an x-ray frequency of $\omega\approx 25$\,keV for reasons of computational feasibility. Modern x-ray laser facilities operate, though, at much lower values of $\xi\sim 10^{-4}$. However, we emphasize that our results can be transferred to other frequencies, as well. While the main parameter dependencies in the far-below threshold regime are given analytically by Eq.~\eqref{sigma-Schwinger-like}, we have checked that our results in the around-threshold and far-above threshold regimes can be well approximated by taking a phase average over the field-free cross section $\sigma_{\rm vac}(s_p) = \frac{4\pi}{3}\frac{\alpha^2}{s_p}\sqrt{1-\frac{4M^2}{s_p}} \big(1+\frac{2M^2}{s_p}\big)\, \theta(\sqrt{s_p}\!-\!2M)$ \cite{Peskin}, where $\theta(x)$ denotes the Heaviside function, according to
\begin{equation}
\label{average}
\sigma \approx \langle \sigma_{\rm vac}(s(\varphi)) \rangle 
= \frac{1}{2\pi}\int_0^{2\pi}\sigma_{\rm vac}(s(\varphi))\,{\rm d}\varphi \,.
\end{equation}
This is because the cross section in these regimes results from $e^+e^-$ collisions whose energy lies above the threshold in the presence of the laser field. Since the field-modulated collision energy $\sqrt{s(\varphi)}$ depends on $\xi$ but not on $\omega$, the integral in Eq.~\eqref{average} is independent of the field frequency. The approximate formula \eqref{average} loses its accuracy when the far-below threshold regime is closely approached, where the cross section becomes strongly $\omega$-dependent [see Eq.~\eqref{sigma-Schwinger-like}].

Moreover, the x-ray scenario considered in our paper can in principle be realized with current technology. To this end, one would let an electron and positron beam of very high energy $p_\pm^0\sim 100$\,GeV cross each other under a very small angle $\vartheta\sim0.1^\circ$ and irradiate the crossing point with a counterpropagating high-intensity optical laser pulse. Viewing this setup from the frame of reference, where the $e^-$ and $e^+$ collide head on, the laser frequency would be Doppler upshifted to the x-ray domain and the scenario considered in our paper would result.

\section{Conclusion}
Production of $\mu^+\mu^-$ pairs in near-threshold $e^+e^-$ collisions assisted by a strong laser field has been studied. While the process $e^+e^-\to\mu^+\mu^-$ in vacuum has a sharp threshold at $\sqrt{s_p}=2M$, below which the cross section vanishes, the threshold condition is softened by the presence of a laser field and the process can, in principle, occur at any value of $\sqrt{s_p}$. Three different interaction regimes have been identified which depend on the laser intensity parameter $\xi$ and the difference $\Delta_p=2M-\sqrt{s_p}$ between the field-free threshold energy and the incident $e^+e^-$ center-of-mass energy: 

When $\Delta_p>2m\xi$, meaning that the collision occurs rather far below the vacuum threshold, the muons are produced via a quantum tunneling process. It features an exponentially suppressed cross section, which shares similarities with the Schwinger effect, but involves some interesting, process-specific parameter dependencies. Conversely, when the $e^+e^-$ collision occurs sufficiently far above threshold with $-\Delta_p>2m\xi$, the cross section approaches its vacuum value.
Interestingly, there is also an intermediate regime around the $2M$ threshold, where the laser field plays an ambivalent role: on the one side, it facilitates a sizeable $\mu^+\mu^-$ production by lifting incident collision energies of $\sqrt{s_p}<2M$ via field acceleration above the threshold; on the other side, the cross section in collisions with $\sqrt{s_p}>2M$ is decreased in the presence of the laser field due to the decelerating field phases.

%%%%%%%%%%%%%%%%%%%%%%%%%%%%%%%%%%%%%%%%%%%%%%%%%%%%%%%%%%%%%%%%%%%%%%%%%%
\section*{Acknowledgment}
%%%%%%%%%%%%%%%%%%%%%%%%%%%%%%%%%%%%%%%%%%%%%%%%%%%%%%%%%%%%%%%%%%%%%%%%%%
This study has been funded by the Deutsche Forschungsgemeinschaft (DFG) 
under Grant No.~392856280 within the Research Unit FOR 2783/2.

%%%%%%%%%%%%%%%%%%%%%%%%%%%%%%%%%%%%%%%%%%%%%%%%%%%%%%%%%%%%%%%%%%%

%%%%%%%%%%%%%%%%%%%%%%%%%%%%%%%%%%%%%%%%%%%%%%%
\end{document}